\documentclass[journal,twoside,web]{ieeecolor}
\usepackage{lcsys}

\usepackage{bm}
\usepackage{enumerate}
\usepackage{commath}
\usepackage{graphicx}
\graphicspath{{Pictures/}}
\usepackage{amsmath}
\usepackage{subcaption}
\usepackage{breqn}
\usepackage{cite}
\usepackage[table]{xcolor}
\usepackage{booktabs}
\usepackage{empheq}
\usepackage{amsfonts}
\usepackage{amssymb}

\usepackage{amsthm}
\usepackage{hyperref}
\usepackage{xcolor}
\usepackage{mathrsfs}
\usepackage{accents}
\usepackage{thmtools}
\usepackage{thm-restate}
\usepackage{float}
\usepackage{xcolor}
\usepackage{comment}
\usepackage[normalem]{ulem}
\usepackage{algpseudocode}
\usepackage{algorithm}

\theoremstyle{plain}
\newtheorem{lemma}{Lemma}
\newtheorem{definition}{Definition}

\newtheorem{property}{Property}
\newtheorem{cor}{Corollary}
\newtheorem{proposition}{Proposition}
\newtheorem{remark}{Remark}
\newtheorem{example}{Example}

\newtheorem*{problem*}{Problem}
\newtheorem*{theorem*}{Theorem}
\newtheorem{assumption*}{Assumption}

\declaretheorem[name=Theorem]{thm}

\newcommand{\redtext}[1]{{\color{red}#1}}
\newcommand{\bluetext}[1]{{\color{black}#1}}
\newcommand{\rev}[1]{{\color{black}#1}}

\newcommand{\myvar}[1]{\bm{#1}}
\newcommand{\myvarfrak}[1]{\bm{\mathfrak{#1}}}
\newcommand{\barvar}[1]{\bar{\bm{#1}}}

\newcommand{\myvardot}[1]{\dot{\myvar{#1}}}

\newcommand{\myset}[1]{\mathscr{#1}}
\newcommand{\mysetbound}[1]{\partial \mathscr{#1}}
\newcommand{\mysetint}[1]{\text{Int}(\mathscr{#1})}

\newcommand{\bracketmat}[2]{ \left[ \begin{array}{#1} #2 \end{array} \right] }

\title{A Robust, Multiple Control Barrier Function Framework for Input Constrained Systems
}

\author{Wenceslao Shaw Cortez, Xiao Tan, and Dimos V. Dimarogonas 
\thanks{This work was supported by the Swedish Research Council (VR),
the Swedish Foundation for Strategic Research (SSF), the Knut and Alice
Wallenberg Foundation (KAW), the H2020-EU Research  and Innovation Programme under the GA No. 101016906 (CANOPIES), and the H2020 ERC Consolidator Grant LEAFHOUND.
The authors are with the School of EECS, Royal Institute of Technology (KTH), 100 44 Stockholm, Sweden (Email: 
        {\tt\small wencsc, xiaotan, dimos@kth.se}).}
}

\begin{document}

\maketitle
\thispagestyle{empty}
\pagestyle{empty}

\begin{abstract}
  We propose a novel (Type-II) zeroing control barrier function (ZCBF) for safety-critical control, which \rev{generalizes} the original ZCBF approach. Our method allows for applications to a larger class of systems (e.g. passivity-based) \rev{while still ensuring robustness}, for which the construction of \rev{conventional} ZCBFs is difficult. \rev{We also propose a \rev{locally Lipschitz continuous} control law that handles multiple ZCBFs, while respecting input constraints, which is not currently possible with existing ZCBF methods.} We apply the proposed concept for unicycle navigation in an obstacle-rich environment.
\end{abstract}

\section{Introduction}

\rev{Recently}, safety-critical control has been associated with zeroing control barrier functions (ZCBFs) \cite{Ames2019}. A safety-critical controller renders a desired constraint set forward invariant for a nonlinear system. \rev{Forward invariance} of the superlevel sets of a ZCBF is ensured if the derivative of the ZCBF is non-negative on the constraint boundary. A minimum-norm quadratic program (QP)-based control law was proposed to enforce this non-negativity condition \cite{Ames2019}. ZCBFs were also associated with asymptotic stabilization to the (compact) constraint set \cite{Xu2015a}\rev{, which provides} robustness to model perturbations/disturbances. Novel developments have addressed input constraints \cite{ShawCortez2020a, Ames2021}, multiple ZCBFs \cite{xu2018constrained}, sampled-data control \cite{ShawCortez2021a,Breeden2022}, self-triggering \cite{Yang2019}, safety and stability \cite{jankovic2018robust}, input-to-state safety \cite{Alan2021}, adaptive/data-driven methods \cite{Lopez2021}, and high-order ZCBFs \cite{Tan2021a, xiao2019control}.

However, there exist limitations in the ZCBF formulation. The ZCBF definition is restrictive \rev{because it requires the ZCBF to strictly decrease} outside of the constraint set to ensure robustness, however passivity-based methods tend to rely on LaSalle's principle which is associated with a non-increasing barrier function. Furthermore, robustness results should be applicable to non-compact sets (with compact boundary), \rev{which occur in} obstacle avoidance scenarios.

Also, the predominant ZCBF controller in the literature is the minimum-norm QP control law. \rev{There exists no guarantee that the minimum-norm QP for a single ZCBF with input constraints is locally Lipschitz continuous \cite{Ames2017}, which means that any guarantees of safety may be nullified. Furthermore,} for multiple ZCBFs, the main approach has been to stack new ZCBF constraints into the QP. However this poses a problem since first, one must ensure that all the ZCBFs are non-conflicting, and second, the aggregation of multiple ZCBF QP constraints will eventually lead to an over-constrained QP. This results in several issues. First, \rev{there is once again no guarantee of} Lipschitz continuity of such QP-based controllers \rev{and so no guarantees of safety can be provided}. This problem could be overcome using sampled-data based ZCBF methods \cite{ShawCortez2021a}, however the fact remains that as the number of ZCBFs grows, the QP will become too large and inefficient for implementation. Finally, this issue becomes exacerbated by considering input constraints \rev{\emph{and}} the multiple ZCBF constraints. There are few methods in the literature that can handle multiple ZCBFs and input constraints simultaneously. In both \cite{Ames2021} and \cite{xu2018constrained}, the authors admit that handling input constraints and multiple ZCBFs simultaneously is a focus of future work. In \cite{ShawCortez2021b}, multiple ZCBFs for input constrained systems are handled, but significant knowledge of the model including Lipschitz constants and bounds on the dynamics is required. Due to these limitations, we seek a novel formulation that allows for a) a more general ZCBF definition that can be applied to passivity-based methods and robustness of non-compact constraint sets b) facilitation of multiple ZCBF design, and c) incorporation of input constraints in light of a) and b).

We propose a novel, robust ZCBF framework for multiple ZCBFs that can handle input constraints. Our first main contribution is the development of a \rev{robust} ``Type-II" ZCBF that relaxes the requirements of the original ZCBF \rev{and can be applied to more general systems}. We propose a mixed-initiative controller \rev{that ensures safety, while respecting input constraints}. Our second contribution is the extension to multiple Type-II ZCBFs \rev{with non-intersecting} set boundaries, \rev{while still respecting input constraints}. We apply the proposed formulation to the unicycle system and present numerical results to demonstrate the proposed approach.

\textit{Notation}: The Lie derivatives of a function $h(\myvar{x})$ for the system $\myvardot{x} = \myvarfrak{f}(\myvar{x}) + \myvarfrak{g}(\myvar{x}) \myvar{u}$ are $L_{\mathfrak{f}} h = \frac{\partial h}{ \partial \myvar{x}} \myvarfrak{f}(\myvar{x}) $ and $L_{\mathfrak{g}} h = \frac{\partial h}{ \partial \myvar{x}} \myvarfrak{g}(\myvar{x})$, resp. The interior and boundary of a set $\myset{A}$ are $\text{Int}(\myset{A})$ and $\partial \myset{A}$, resp. The distance from  $\myvar{x}$ to a set $\myset{A}\subset \mathbb{R}^n$ is $\| \myvar{x} \|_{\myset{A}} := \inf_{\myvar{w}\in \myset{A}} \| \myvar{x} - \myvar{w} \| $. A uniformly continuous function $\myvar{x}:\mathbb{R}_{\geq 0} \to \mathbb{R}^n$ asymptotically approaches a set $\myset{X} \subset \mathbb{R}^n$, if as $t \to \infty$, for each $\varepsilon \in \mathbb{R}_{>0}$, $\exists T \in \mathbb{R}_{>0}$, such that $\|\myvar{x}(t)\|_{\myset{X}} < \varepsilon \ \forall t\geq T$. A continuous function $\alpha:\mathbb{R} \to \mathbb{R}$ is an extended class $\mathcal{K}$ function if it is strictly increasing and $\alpha(0) = 0$. For a given set $\Omega$ and system $\myvardot{x} = \myvar{f}(\myvar{x})$, no solution of the system can stay identically in $\Omega$ if for some $\tau_1 \in \mathbb{R}_{\geq 0}$ for which $\myvar{x}(\tau_1) \in \Omega$, there exists a $\tau_2 \in \mathbb{R}_{\geq0}$, $\tau_2 > \tau_1$, such that $\myvar{x}(\tau_2) \notin \Omega$.

\section{Background}

\subsection{Zeroing Control Barrier Functions}

\rev{Consider} the nonlinear affine system: 
\begin{equation}\label{eq:nonlinear affine dynamics}
 \myvardot{x} = \myvar{f}(\myvar{x}) + \myvar{g}(\myvar{x}) \myvar{u},
\end{equation}
 where $\myvar{f}: \myset{X} \to \mathbb{R}^n$ and $\myvar{g}: \myset{X} \to \mathbb{R}^{n\times m}$ are locally Lipschitz continuous functions on their domain $\myset{X} \subseteq \mathbb{R}^n$, $\myvar{u}: \myset{X} \to \myset{U} \subseteq \mathbb{R}^m$ is the control input, and $\myvar{x}(t, \myvar{x}_0) \in \myset{X}$ is the state trajectory at $t$ starting at $\myvar{x}_0 \in \myset{X}$, which with abuse of notation we denote $\myvar{x}(t)$.

Let $h(\myvar{x}): \myset{X} \to \mathbb{R}$ be a continuously differentiable function, and let the associated constraint set be defined by:
\begin{equation} \label{eq:constraint set general}
\myset{C} = \{\myvar{x} \in \myset{X}: h(\myvar{x}) \geq 0\}.
\end{equation} 

\begin{definition}\label{def:safe}
Consider the system \eqref{eq:nonlinear affine dynamics} under a given control law $\myvar{u}$ and the maximal interval of existence $\myset{I} \subseteq \mathbb{R}_{\geq 0}$ of the solution $\myvar{x}(t)$. The system \eqref{eq:nonlinear affine dynamics} with respect to a given closed set $\myset{C} \subset \myset{X}$ is \textit{safe} if $\myvar{x}(0) \in \myset{C}$, then $\myvar{x}(t) \in \myset{C}$ for all $t\in \myset{I}$.
\end{definition}

Constraint satisfaction is ensured by showing that the system states are always directed into the constraint set. If there exists a control input to satisfy this condition, then $h$ is considered a ZCBF:
\begin{definition}[\cite{Ames2019}]\label{def:zcbf}:
Let the set $\myset{C} \subset \myset{D} \subset \myset{X}$ be the superlevel set of a continuously differentiable function $h: \myset{D} \to \mathbb{R}$. Then $h$ is a zeroing control barrier function (ZCBF)  if there exists an extended class-$\mathcal{K}$ function $\alpha$ such that for the control system \eqref{eq:nonlinear affine dynamics}:
\begin{align}\label{eq:ZCBF condition}
 \underset{\myvar{u} \in \myset{U}}{\text{sup}} [L_f h (\myvar{x}) + L_g h(\myvar{x}) \myvar{u} + \alpha(h(\myvar{x}))] \geq 0, \forall \myvar{x} \in \myset{D}
\end{align}
\end{definition}

The advantage of checking \eqref{eq:ZCBF condition} over all of $\myset{D}$, for which $h$ may be negative, is to ensure asymptotic stability to the set $\myset{C}$, provided that $\myset{C}$ is compact  \cite{Xu2015a}. One way to implement \bluetext{(multiple) ZCBFs, assuming they are non-conflicting, is by using the following QP-based controller:
\begin{align}\label{eq:zcbf qp}
\myvar{u}^*(\myvar{x}) \hspace{0.1cm} = \hspace{0.1cm} & \underset{\myvar{u} \in \myset{U}} {\text{argmin}}
\hspace{.3cm} \| \myvar{u} -\myvar{u}_{\text{nom}}(\myvar{x},t) \|^2_2  \\
& \text{s.t.} \hspace{.1cm}  L_f h_j(\myvar{x}) + L_g h_j(\myvar{x}) \myvar{u} \geq - \alpha_j(h_j(\myvar{x})), j \in \myset{N} \nonumber
\end{align}
where $\myset{N} = \{1,..,N\}$ for $N \in \mathbb{N}$} and $\myvar{u}_{nom}: \mathbb{R}^n \times \mathbb{R}_{\geq0} \to \mathbb{R}^m$ is any nominal Lipschitz continuous controller that could be, e.g., a stabilizing controller or human input. \rev{When considering \eqref{eq:zcbf qp}, or similar controllers \cite{Ames2017}, with input constraints, there is no guarantee that the control law is locally Lipschitz continuous, even with only a \emph{single} ZCBF \cite{Ames2017}. Local Lipschitz continuity of the controller is required for ensuring safety \cite{Ames2019}. Thus since \eqref{eq:zcbf qp} fails to satisfy the safety conditions, any guarantees of safety may be nullified.}

\section{Type-II ZCBFs}\label{sec:type II ZCBFs}
\subsection{Safety and Robustness}
We expand the concept of ZCBFs for more general applications wherein a function satisfying Definition \ref{def:zcbf} may not exist or is difficult to construct. We propose an alternative ZCBF, referred to as a Type-II ZCBF, that is less restrictive than that of Definition \ref{def:zcbf}. To begin, we define the following properties to replace extended class-$\mathcal{K}$ functions:
\begin{property}\label{prop:typeII}
The  function, $\alpha: \mathbb{R} \to \mathbb{R}$ is continuous and the restriction of $\alpha$ to $\mathbb{R}_{\geq 0}$ is of class-$\mathcal{K}$.
\end{property}

\begin{property}\label{prop:typeII robust}
The function $\alpha: \mathbb{R} \to \mathbb{R}$ satisfies: $\alpha(h) \leq 0, \forall h< 0$.
\end{property}

Next, we specify that we do \emph{not} require the ZCBF condition to hold in a superlevel set $\myset{D}$ containing all of $\myset{C}$. Instead, we only require a designer to check that the ZCBF condition holds in a neighborhood around $ \mysetbound{C}$ defined as follows:
\begin{equation}\label{eq:constraint annulus}
    \myset{A} = \{ \myvar{x} \in \myset{X}: h(\myvar{x}) \in [-b, a ]  \}
\end{equation}
for some $a,b \in \mathbb{R}_{> 0}$.

\begin{definition}[Type-II ZCBF]\label{def:typeII ZCBF}
Given the set $\myset{C}$ defined by \eqref{eq:constraint set general} for a continuously differentiable function $h: \myset{X} \to \mathbb{R}$, the function $h$ is called a \emph{Type-II ZCBF} with respect to the set $\myset{C}$ defined in \eqref{eq:constraint set general} if there exists a function $\alpha$ satisfying Property \ref{prop:typeII} and a set $\myset{A}$ defined in \eqref{eq:constraint annulus} such that the following holds:
\begin{align}\label{eq:typeII ZCBF condition}
 \underset{\myvar{u} \in \myset{U}}{\text{sup}} [L_f h (\myvar{x}) + L_g h(\myvar{x}) \myvar{u} + \alpha(h(\myvar{x}))] \geq 0, \forall \myvar{x} \in \myset{A}
\end{align}
\end{definition}

If $h$ is a Type-II ZCBF, then set of control inputs satisfying $\eqref{eq:typeII ZCBF condition}$ is: \rev{$\myset{S}(\myvar{x}) = \{ \myvar{u} \in \myset{U}:\text{ if }\myvar{x} \in \myset{A}, \text{ then }\eqref{eq:typeII ZCBF condition} \text{ holds } \}$.}

\begin{thm}\label{thm:typeII safety AS}
Consider the system \eqref{eq:nonlinear affine dynamics} and the set $\myset{C} \subset \myset{X}$ from \eqref{eq:constraint set general} for the continuously differentiable function $h: \myset{X} \to \mathbb{R}$. Suppose $h$ is a \textit{Type-II ZCBF} for a given $\alpha$ and $\myset{A}$ defined by \eqref{eq:constraint annulus}, and $\nabla h(\myvar{x} ) \neq 0$ for all $ x\in \mysetbound{C}$. 
\begin{enumerate}[(i)]
    \item If there exists a locally Lipschitz continuous $\myvar{u}: \myvar{x} \in \myset{X} \mapsto  \myset{S}(\myvar{x})$, then the closed-loop system is safe with respect to $\myset{C}$.
    \item In addition to (i), suppose $\myset{A}$ is compact, $\alpha$ satisfies Property \ref{prop:typeII robust}, $\myvar{x}(t) \in \myset{X}$ is bounded for all $t\geq 0$, and let $\myset{D} = \myset{C} \cup \myset{A}$ be a connected set. If no solution of the closed-loop system can stay identically in the set $\Omega := \{ \myvar{x} \in \myset{D}\setminus \myset{C}: \dot{h}(\myvar{x}) = 0\}$, then $\myset{C}$ is an asymptotically stable set.
\end{enumerate}  
\end{thm}
\begin{proof}
(i) Since $h$ is a Type-II ZCBF, there exists a control $\myvar{u} \in \myset{U}$ satisfying \eqref{eq:typeII ZCBF condition} and so $\myset{S}(\myvar{x})$ is non-empty. The closed-loop dynamics are locally Lipschitz on the open set $\myset{C} \cup \mysetint{A}$, such that the solution of the closed-loop system is uniquely defined on $\myset{I} = [0, \tau)$ for some $\tau \in \mathbb{R}_{>0}$. Since $h$ is a Type-II ZCBF,  $\dot{h} \geq 0$ holds on the boundary of $\myset{C}$, which is equivalent to condition (1) of Brezis' Theorem (Theorem 1 of \cite{Brezis1970}). Thus Brezis' Theorem ensures $\myvar{x}(t) \in \myset{C}$ for all $t \in \myset{I}$ and the closed-loop system is safe with respect to $\myset{C}.$ 

(ii) Since $h$ is a Type-II ZCBF with an $\alpha$ satisfying Property \ref{prop:typeII robust}, and $\myvar{u} \in \myset{S}(\myvar{x})$, then $\dot{h} \geq - \alpha(h) \geq 0$ for all for all $\myvar{x} \in \myset{A}$. Let a) $V = -h$ if $h < -1$, b) $V = h^3 + 2h^2$ if $h \in [-1, 0]$, and c) $V = 0$ if $h > 0$. It is clear that $V$ is continuously differentiable. Furthermore, $\dot{V} \leq 0$ for all $\myvar{x} \in \myset{D}$ since $\dot{h} \geq 0 $ in $\myset{A}$ and $3h^2 + 4h <= 0$ for $h \in [-1, 0]$. Also, Brezis' Theorem ensures the closed-loop system is safe with respect to $\myset{D}$ for all $t \geq 0$ because the closed-loop system is locally Lipschitz on $\myset{X} \supset \myset{D}$, $\dot{h} \geq 0$ on the boundary of $\myset{D}$, and $\myvar{x}(t)$ is defined for all $t \geq 0$.

Since $\myvar{x}(t) \in \myset{X}$ and $\myvar{x}(t)$ is bounded in $\myset{X}$, Lemma 4.1 of \cite{Khalil2002} ensures that $\myvar{x}(t) \to L^+$ as $t \to \infty$, where $L^+$ is the positive limit set of the closed-loop system. Furthermore, since $\myset{A}$ is compact and $V$ is continuous, $V$ is lower bounded on $\myset{A}$ by $V(\myvar{x}) = 0$. We see that $V(\myvar{x}(t))$ is a monotonically decreasing function of $t$. We note that since $\myset{A}$ is compact, if $V(\myvar{x}(t))$  reaches zero, it must reach zero in $\myset{A}$. Let $\bar{\Omega}:= \{ \myvar{x} \in \myset{D}: \dot{V}(\myvar{x}) = 0\}$ (note that $\dot{V} = 0$ is equivalent to $\dot{h} = 0$ in $\myset{D} \setminus \myset{C}$), for which $L^+$ is a subset of the largest invariant set in $\bar{\Omega}$. From the proof of Theorem 4.4 of \cite{Khalil2002}, $\myvar{x}(t)$ approaches $L^+ \subset \bar{\Omega}$ as $t \to \infty$. Furthermore, since no solution can stay identically in $\Omega$, then $L^+ \cap \Omega = \emptyset$ and so $\myvar{x}(t)$ approaches $\bar{\Omega}\setminus \Omega \subset \myset{C}$. Thus $\myset{C}$ is an attractor \cite{El-Hawwary2007}.

 
 Since $\myset{A}$ is compact, $\|\myvar{x}\|_{\myset{C}}, V(\myvar{x}) > 0$ when $\myvar{x} \in \myset{D}\setminus \myset{C}$, and $\|\myvar{x}\|_{\myset{C}}, V(\myvar{x}) = 0$ when $\myvar{x} \in \myset{C}$ we can always find class-$\mathcal{K}$ functions $\alpha, \beta$ such that $\alpha(\|\myvar{x}\|_{\myset{C}}) \leq V(\myvar{x}) \leq \beta(\|\myvar{x}\|_{\myset{C}} )$ for all $\myvar{x} \in \myset{A}$. Since $\dot{V} \leq 0$, $\myset{C}$ is uniformly stable (see e.g. Corollary 1.7.5 of \cite{Bathia1967}), which in addition to being an attractor implies that $\myset{C}$ is asymptotically stable.
\end{proof}

\begin{cor}\label{cor:AS compact set}
Suppose the conditions of Theorem \ref{thm:typeII safety AS} hold up to and including (i), $\alpha$ satisfies Property \ref{prop:typeII robust}, and let $\myset{D} = \myset{C} \cup \myset{A}$ be a compact, connected set. If no solution of the closed-loop system can stay identically in the set $\Omega := \{ \myvar{x} \in \myset{D}\setminus \myset{C}: \dot{h}(\myvar{x}) = 0\}$, then $\myset{C}$ is an asymptotically stable set.
\end{cor}
\begin{proof}
 \rev{Similar to Theorem \ref{thm:typeII safety AS} and omitted for brevity.}
\end{proof}

The results of Theorem \ref{thm:typeII safety AS} and Corollary \ref{cor:AS compact set} generalize the ZCBF results of \cite{Ames2019}, and ZCBFs are in fact a subset of Type-II ZCBFs. The Type-II ZCBF condition \eqref{eq:typeII ZCBF condition} is only required in a neighborhood around $\mysetbound{C}$, which allows for a new control design for handling multiple ZCBFs under input constraints as will be shown in the following section. Regarding robustness, the original ZCBFs require $h$ to strictly decrease outside of $\myset{C}$ for set asymptotic stability. The Type-II ZCBFs only require $h$ to be non-increasing outside of $\myset{C}$ because LaSalle's principle is exploited to facilitate the ZCBF design. The condition that no solution can stay identically in  $\Omega$ is similar to zero-state observability \cite{Khalil2002} and is trivially satisfied if $h$ is a ZCBF from Definition \ref{def:zcbf}. The proposed method can be applied to passive systems (see Example \ref{ex:passivity1} and Section \ref{sec:unicycle}), for which our results can be extended to non-compact $\myset{A}$ \emph{and} $\myset{C}$ \cite{El-Hawwary2007}, \cite{El-Hawwary2008}.

\begin{example}\label{ex:passivity1}
Consider the mechanical system: $\myvardot{q} = \myvar{v}$, $\myvardot{v}$ $=$ $M^{-1}(\myvar{q}) \left( -C(\myvar{q}, \myvar{v})\myvar{v} - \myvar{g}(\myvar{q}) - F \myvar{v} + \myvar{u}\right)$, where $M: \myset{M} \to \mathbb{R}^{n \times n}$ is the positive-definite inertia matrix, $C: \myset{M} \times \mathbb{R}^n \to \mathbb{R}^{n\times n}$ is the Coriolis and centrifugal term, $\myvar{g}: \myset{M} \to \mathbb{R}^n$ is the gravity torque, and $F \in \mathbb{R}^{n\times n}$ is a positive-definite damping matrix. \rev{The state is $\myvar{x}(t) = (\myvar{q}(t), \myvar{v}(t)) \in \mathbb{R}^{2n}$. For $\myvar{u} = \myvar{g} + \myvar{\mu}$, the system is passive with respect to $\myvar{\mu}$ and output $\myvar{y} = \myvar{v}$ with storage function $S(\myvar{x}) = \frac{1}{2} \myvar{v}^TM(\myvar{q}) \myvar{v}$ such that $\dot{S} \leq \myvar{\mu}^T \myvar{y}$.} \rev{Consider the} constraint set $\myset{Q} = \{ \myvar{q} \in \myset{M}: c(\myvar{q}) \geq 0 \}$ for the continuously differentiable function $c: \myset{M} \to \mathbb{R}_{\geq 0}$. We define the Type-II ZCBF candidate as $h(\myvar{x}) = \rev{k_h} c(\myvar{q}) - \rev{S(\myvar{x})}$, wherein \rev{$\dot{h} = k_h \nabla c^T \myvar{v} - \dot{S} \geq k_h \nabla c^T \myvar{v} - \myvar{\mu}^T \myvar{y} \geq 0$} on any $\myset{A}$ from \eqref{eq:constraint annulus} with $\myvar{u}_{s} = \myvar{g} + k_h \nabla c$, \rev{($\myvar{\mu} = k_h \nabla c$)}. Thus $h$ is a Type-II ZCBF \rev{with $\alpha(h) = \bar{\alpha}(h) \text{ if } h \geq 0$, and $\alpha(h) = 0 \text{ if } h< 0 $, for any class-$\mathcal{K}$ function $\bar{\alpha}$. We satisfy input constraints, assuming $\myvar{g}(\myvar{q}) \in \mysetint{U}$, by defining $k_h \leq \max_{k \in \mathbb{R}_{>0}} k \text{ s.t. } \myvar{g}(\myvar{q}) + k \nabla c(\myvar{q}) \in \myset{U} \forall \myvar{x} \in \myset{A}$. Such a $k$ always exists if $\myset{U}$ is closed and $\myset{A}$ is compact since $\nabla c$ is a continuous function.} In \cite{ShawCortez2021}, we required $\myvar{g} \neq \myvar{u}$ in $\myset{D} \setminus \myset{C}$, which is equivalent to ensuring no solution can stay identically in $\Omega$ from Theorem \ref{thm:typeII safety AS} and Corollary \ref{cor:AS compact set}. In \cite{ShawCortez2021} we provided guarantees of set attractiveness for $\myset{C}$, whereas here we extend those results to asymptotic stability of the safe set for (non-)compact sets. \rev{The approach presented here, i.e., using the storage function for constructing a Type-II ZCBF, can be applied to other systems, including, e.g., the double integrator.}
\end{example}

\subsection{Mixed-Initiative ZCBF Controller}

Here we present a control law to implement the Type-II ZCBFs. We introduce the mixed-intiative controller for a given nominal control $\myvar{u}_{nom}:\myset{X} \to \myset{U}$ and locally Lipschitz continuous safety controller $\myvar{u}_s: \myvar{x} \in \myset{X} \mapsto \myset{S}(\myvar{x}) \subset \myset{U}$ as:
\begin{equation}\label{eq:mic ZCBF}
    \myvar{u}^* = \left(1 - \phi_a(h(\myvar{x})\right) \myvar{u}_s + \phi_a(h) \myvar{u}_{nom},
\end{equation}
where $\phi_{a}: \mathbb{R} \to [0, 1]$ is defined by:
\begin{align} \label{eq:phi}
    \phi_{a}(h) = \begin{cases}
    1, \text{ if } h > a \\
    \kappa(h), \text{ if } h \in [0, a] \\
    0, \text{ if } h < 0
    \end{cases}
\end{align}
where $\kappa: \mathbb{R} \to [0,1]$ is any locally Lipschitz continuous function that satisfies $\kappa(0) = 0$ and $\kappa(a) = 1$, for $a$ from \eqref{eq:constraint annulus}. \rev{The choice of $\kappa$ dictates how aggressive the controller is as the system approaches $\mysetbound{C}$. We also see the effect of $a, b$ in \eqref{eq:mic ZCBF} and \eqref{eq:typeII ZCBF condition}. Ideally, $a$ should be small to reduce the interference of $\myvar{u}_{nom}$, but if $a$ is too small, the controller may be sensitive to measurement noise. The larger $b$ is (and hence larger $\myset{A}$), the larger the region of attraction for $\myset{C}$ is to handle larger disturbances. However, a larger $\myset{A}$ may require more control authority}. Also, \eqref{eq:mic ZCBF} is a point-wise convex combination of $\myvar{u}_{s}$ and $\myvar{u}_{nom}$ such that if $\myset{U}$ is convex, then $\myvar{u}^* \in \myset{U}$, and thus input constraints are satisfied in a straightforward fashion, as shown in the following theorem:

\begin{thm}\label{thm: type II ZCBF mic}
Consider the system \eqref{eq:nonlinear affine dynamics} and the set $\myset{C} \subset \myset{X}$ from \eqref{eq:constraint set general} for the continuously differentiable function $h: \myset{X} \to \mathbb{R}$. If $h$ is a Type-II ZCBF, $\myset{U}$ is convex, and $\myvar{u}_s: \myset{X} \to \myset{S}(\myvar{x})$ and $\myvar{u}_{nom}: \myset{X} \to \myset{U}$ are locally Lipschitz continuous, then $\myvar{u}^*(\myvar{x}) \in \myset{U}$ for all $\myvar{x} \in \myset{X}$ and $\myvar{u}^*$ is locally Lipschitz continuous. Furthermore, $\myvar{u}^*$ in closed-loop with \eqref{eq:nonlinear affine dynamics} renders the system safe with respect to $\myset{C}$.
\end{thm}
\begin{proof}
Local Lipschitz continuity of $\myvar{u}^*$ follows directly from \eqref{eq:mic ZCBF}. It is clear that for every $\myvar{x} \in \myset{X}$, $\myvar{u}^*$ is a convex combination of $\myvar{u}_s$ and $\myvar{u}_{nom}$, which are both elements of the convex set $\myset{U}$. Thus $\myvar{u}^* \in \myset{U}$. By construction of $\myvar{u}^*$, $\myvar{u}^* = \myvar{u}_s$ on the boundary $\mysetbound{C}$, and since $h$ is a Type-II ZCBF then from Theorem \ref{thm:typeII safety AS} the system \eqref{eq:nonlinear affine dynamics} is safe.
\end{proof}

\rev{
\begin{remark}\label{rem:QP type-II}
One can construct $\myvar{u}_s$ for a given Type-II ZCBF $h$ as: a) if $\myvar{x} \in \myset{A}$, $\myvar{u}_s = \min_{\myvar{u} \in \mathbb{R}^m} \|\myvar{u}\|_2^2$ s.t. $L_fh + L_gh \myvar{u} \geq - \alpha(h)$ and b) if $\myvar{x} \notin \myset{A}$, $\myvar{u}_s = 0$. If $L_gh \neq 0$ on $\myset{A}$ and $\nabla h(\myvar{x})$ and $\alpha(h)$ are locally Lipschitz, then it is clear from \cite{hager1979} that $\myvar{u}_s$ is locally Lipschitz on $\mysetint{A}$ and that when implemented in $\myvar{u}^*$ from $\eqref{eq:mic ZCBF}$, $\myvar{u}^*$ is also locally Lipschitz since whenever $\myvar{x} \notin \mysetint{A}$, then $\myvar{u}^* = \myvar{u}_{nom}$. Furthermore, for $\myset{U} = \{ \myvar{u} \in \mathbb{R}^m: \|\myvar{u}\|_2 \leq \theta \}$, $\theta \in \mathbb{R}_{\geq 0}$, since Definition \ref{def:typeII ZCBF} ensures there exists a $\myvar{u} \in \myset{U}$ to satisfy \eqref{eq:typeII ZCBF condition} and $\myvar{u}_s$ is the minimum-norm control to enforce \eqref{eq:typeII ZCBF condition}, then $\myvar{u}_s \in \myset{U}$ and the results from Theorem \ref{thm: type II ZCBF mic} hold. We note that one must check $\Omega$ to ensure robustness if $\alpha$ is not an extended class-$\mathcal{K}$ function. Alternatively, the approach presented in Example \ref{ex:passivity1} provides another means of constructing $\myvar{u}_s$.
\end{remark}
}

\subsection{Multiple Type-II ZCBFs}

Here we address N Type-II ZCBFs, while respecting input constraints. Consider $h_i: \myset{X} \to \mathbb{R}$ for $i \in \myset{N}:= \{1,,,,.N\}$ for $a_i, b_i$ from \eqref{eq:constraint annulus}, $\alpha_i$ from \eqref{eq:typeII ZCBF condition}, and let $\phi_{a_i}$ denote \eqref{eq:phi} for constraint $i$. We emphasize that each $a_i, b_i, \alpha_i$ need not be the same for all $i \in \myset{N}$ and so each Type-II ZCBF $h_i$ can be designed independently. We define the associated sets for each Type-II ZCBF as follows, for $i \in \myset{N}$:
\begin{align}\label{eq:safe set i}
    \myset{C}^i = \{ \myvar{x} \in \myset{X}: h_i(\myvar{x}) \geq 0\}, 
\end{align}
\begin{equation}\label{eq:constraint annulus i}
    \myset{A}^i = \{ \myvar{x} \in \myset{X}: h_i(\myvar{x}) \in [-b_i, a_i ] \},
\end{equation} 
\begin{align}\label{eq:typeII ZCBF condition multi}
 \underset{\myvar{u} \in \myset{U}}{\text{sup}} [L_f h_i (\myvar{x}) + L_g h_i(\myvar{x}) \myvar{u} + \alpha_i(h_i(\myvar{x}))] \geq 0, \forall \myvar{x} \in \myset{A}^i,
\end{align}
\rev{and $\myset{S}^i(\myvar{x})$ $=$ $\{ \myvar{u} \in \myset{U}:$ $\text{if }\myvar{x} \in \myset{A}^i,$ $\text{then } \eqref{eq:typeII ZCBF condition multi}$ $\text{holds for }h_i  \}$}. 

For multiple Type-II ZCBFs, if $\myset{A}^i$ for each $i \in \myset{N}$ do not overlap, then whenever the state enters any $\myset{A}^i$, we can implement \eqref{eq:mic ZCBF} for the associated $h_i$ and render $\myset{C}^i$ forward invariant. This provides a straightforward way of independently addressing multiple ZCBFs. In this letter we consider non-overlapping Type-II ZCBFs, and will address the over-lapping case in future work. We define the input constraint satisfying, multiple ZCBF controller as:
\begin{align}\label{eq:mic ZCBF multiple}
    \myvar{u}^*(\myvar{x}) = \left(1 - \bar{\phi}(\myvar{x})\right) \barvar{u}_s(\myvar{x}) + \bar{\phi}(\myvar{x}) \myvar{u}_{nom}(\myvar{x})
\end{align}
where \rev{$\bar{\phi}(\myvar{x}) =\phi_{a_i}(h_i(\myvar{x}))$ if $\myvar{x} \in \myset{A}^i, i \in \myset{N}$ and $\bar{\phi}(\myvar{x}) =1$ otherwise, $\barvar{u}_s(\myvar{x}) =\myvar{u}_{s_i}(\myvar{x})$ if $\myvar{x} \in \myset{A}^i, i \in \myset{N}$ and $\barvar{u}_s(\myvar{x}) = 0$ otherwise.}

\begin{thm}\label{thm:Multiple ZCBF MIC}
Given $N$ continuously differentiable functions $h_i: \myset{X} \to \mathbb{R}$, $i \in \myset{N}$ for the system \eqref{eq:nonlinear affine dynamics}, suppose that $\myvar{u}_{nom}: \myset{X} \to \myset{U}$ is locally Lipschitz continuous. If each $h_i$ is a Type-II ZCBF with associated $\myvar{u}_{s_i}: \myvar{x}\in \myset{X} \mapsto \myset{S}^i(\myvar{x})$, and if for any $j,k \in \myset{N}$, $j \neq k$, $\myset{A}^j \cap \myset{A}^k = \emptyset$, then $\myvar{u}^*$ defined by \eqref{eq:mic ZCBF multiple} implemented in closed-loop with \eqref{eq:nonlinear affine dynamics} ensures that:
\begin{enumerate}[(i)]
    \item If $\myvar{x}(0) \in \bigcap_{i \in \myset{N}}\myset{C}^i$, then the system \eqref{eq:nonlinear affine dynamics} is safe with respect to each $\myset{C}^i$, $i \in \myset{N}$.
    \item If $\myset{U}$ is convex, then $\myvar{u}^*(\myvar{x}) \in \myset{U}$ for all $\myvar{x} \in \myset{X}$.
\end{enumerate}
\end{thm}
\begin{proof}
For any $\myvar{x} \in \myset{X}$, since for any $j,k \in \myset{N}$, $j \neq k$, $\myset{A}^j \cap \myset{A}^k = \emptyset$ either a) there exists a unique $i$ for which $\myvar{x} \in \myset{A}^i$ or b) $\myvar{x} \notin \myset{A}^i$ for any $i \in \myset{N}$. Furthermore, since each $\phi_{a_i}(h_i) = 1$, for $h_i \geq a_i$, $\bar{\phi}$ is well-defined and locally Lipschitz continuous for all $\myvar{x} \in \myset{X}$. Similarly since each $\myvar{u}_{s_i}$ is well-defined on $\myset{A}^i$, for $i \in \myset{N}$, $\barvar{u}_s(\myvar{x})$ is well-defined. Now, $\barvar{u}_s(\myvar{x})$ is locally Lipschitz continuous for $\myvar{x} \in \myset{A}^i$, but may switch when $\myvar{x}$ leaves $\myset{A}^i$. We note however that whenever $\myvar{x} \notin \myset{A}^i$ for any $i \in \myset{N}$, $\bar{\phi} = 0$ such that $\myvar{u}^*= \myvar{u}_{nom}$. Thus $\myvar{u}^*$ is well-defined and locally Lipschitz continuous for all $\myvar{x} \in \myset{X}$ and the proof follows from Theorem \ref{thm: type II ZCBF mic} for each $i \in \myset{N}$.
\end{proof}

The proposed control \eqref{eq:mic ZCBF multiple} has several advantages over the QP formulation \eqref{eq:zcbf qp}. First, \eqref{eq:mic ZCBF multiple} \rev{is guaranteed to be a locally Lipschitz continuous controller that can satisfy multiple Type-II ZCBFs \emph{and} input constraints, which to date is not possible with \eqref{eq:zcbf qp} or similar controllers \cite{Ames2017}}. \rev{Second,} since \eqref{eq:mic ZCBF multiple} only implements $\myvar{u}_{s_i}$ near $\mysetbound{C}^i$, \rev{we know that $\myvar{u} = \myvar{u}_{nom}$ when $\myvar{x} \notin \cup_{i\in \myset{N}} \myset{A}^i$. Thus we know a priori where the nominal control will be implemented which is advantageous for completing tasks, e.g., stabilization. This is not possible with \eqref{eq:zcbf qp}, for which the ZCBF constraint may be active anywhere in $\myset{C}$}. \rev{Third}, as the number of constraints and states grows, the QP \eqref{eq:zcbf qp} becomes excessively large and inefficient to implement, \rev{whereas \eqref{eq:mic ZCBF multiple} scales well}. Finally, \rev{\eqref{eq:mic ZCBF multiple} can still be implemented with QP-based controllers (see Remark \ref{rem:QP type-II}), if optimality is desired, while retaining all the previously mentioned advantages over \eqref{eq:zcbf qp}.}

\section{Application to Unicycle Dynamics}\label{sec:unicycle}

Consider the unicycle dynamics defined by:
\begin{equation}\label{eq:unicycle dynamics}
    \dot{x}_1 = u_p \cos(x_3), \ \dot{x}_2 = u_p \sin(x_3), \ \dot{x}_3 = u_d
\end{equation}
where \rev{$\myvar{z} = (x_1, x_2) \in \mathbb{R}^2$ is the position on the plane, $u_p$, $u_d \in \mathbb{R}$ are the speed and rate of rotation, respectively}, $x_3 \in S^1$ \rev{is the heading angle}, $\myvar{x} = (x_1, x_2, x_3)$, $\myset{X} = \mathbb{R}\times \mathbb{R}\times S^1$, and $\myvar{u} = (u_p, u_d)$ \rev{is the control input}. 

Consider the following \rev{ellipsoid} constraints:
\begin{align}\label{eq:unicycle constraint}
    c_i(\myvar{x}) =  \gamma_i \big( \Delta_i^2 - \frac{1}{2}\|\myvar{e}_i\|_{P_i}^2 \big) 
\end{align}
for some $\gamma_i \in \{-1,1\}$, $\Delta_i \in \mathbb{R}_{\geq 0}$, a symmetric, positive-definite $P_i \in \mathbb{R}^{2\times 2}$, $\myvar{e}_i = \myvar{z} - \myvar{z}_{r_i}$ for \rev{the ellipsoid center} $\myvar{z}_{r_i} \in \mathbb{R}^2$, and $i \in \myset{N}$. The sets that can be defined by \eqref{eq:unicycle constraint} include ellipsoidal obstacles to be avoided as well as ellipsoidal regions the unicycle must stay inside. We propose the following Type-II ZCBFs:
\begin{align}\label{eq:unicycle h}
    h_i(\myvar{x}) =  c_i(\myvar{x}), \ i \in \myset{N}
\end{align}
with safe sets \eqref{eq:safe set i} for $i \in \myset{N}$. \rev{We note that for $\gamma = -1$, $\myset{C}^i$ is non-compact with compact $\myset{A}^i$}. Let $\myvar{u}_{s_i} = (u_{s_{p_i}}, u_{s_{d_i}})$ for:
\begin{subequations}
\label{eq:unicycle control}
\begin{align}
    u_{s_{p_i}} &= \begin{cases} -k_{p_i} \gamma_i c_i \myvar{e}_i^T P_i \bracketmat{cc}{\cos(x_3) & \sin(x_3)}^T, \text{ if } c_i \geq 0 \\
    k_{p_i} \gamma_i c_i \myvar{e}_i^T P_i \bracketmat{cc}{\cos(x_3) & \sin(x_3)}^T , \text{ if } c_i < 0\end{cases}\\
    u_{s_{d_i}} &= - k_{d_i} \myvar{e}_i^T P_i \bracketmat{cc}{- \sin(x_3)& \cos(x_3) }^T
\end{align}
\end{subequations}
where $k_{p_i}, k_{d_i} \in \mathbb{R}_{>0}$ for $i \in \myset{N}$. \rev{We note that \eqref{eq:unicycle control} was motivated by the passivity-based control from \cite{El-Hawwary2008}.} Let:
\begin{equation}\label{eq:unicycle input set}
    \myset{U} = \{ \myvar{u} \in \mathbb{R}^2: | u_p | \leq \bar{u}_p, | u_d | \leq \bar{u}_d \}
\end{equation}
for $\bar{u}_p, \bar{u}_d \in \mathbb{R}_{>0}$.

\begin{proposition}\label{prop:unicycle}
Consider the system \eqref{eq:unicycle dynamics} and the constraint sets $\myset{C}^i$ from \eqref{eq:safe set i} with $h_i$ from \eqref{eq:unicycle h} and $c_i(\myvar{x})$ from \eqref{eq:unicycle constraint} for $i \in \myset{N}$. Suppose that for each given $\myset{A}^i$ from \eqref{eq:constraint annulus i},  $\myset{A}^i \cap \myset{A}^j = \emptyset$ for all $j \in \myset{N}\setminus \{i\}$. Suppose $\myvar{u}_{nom}: \myset{X} \to \myset{U}$ is locally Lipschitz continuous. Consider the system in closed-loop with the control law \eqref{eq:mic ZCBF multiple}, \eqref{eq:unicycle control}. Then:
\begin{enumerate}[(i)]
    \item If $\myvar{x}(0) \in \cap_{i \in \myset{N}} \myset{C}^i$, then  the closed-loop system is safe with respect to each $\myset{C}^i$.
    \item If each $\myset{A}^i$ excludes the point $\myvar{z} = \myvar{z}_{r_i}$ for $i \in \myset{N}$ and either a) there exists an $i \in \myset{N}$ for which $\gamma_i = 1$, or b) $\myvar{u}_{nom}$ is such that $\myvar{x}(t)$ is bounded and well-defined for all $t\geq 0$, then each $\myset{C}^i$ is asymptotically stable for $i \in \myset{N}$.
    \item If $\myset{U}$ is defined by \eqref{eq:unicycle input set} and $k_{p_i}, k_{d_i}$ satisfy:
\begin{equation}\label{eq:unicycle gains}
    k_{p_i} \leq \frac{\bar{u}_p}{  \eta_i \max \{a_i, b_i\}}, 
    k_{d_i} \leq \frac{\bar{u}_d}{ \eta_i  }, \forall i \in \myset{N}
\end{equation}
    where $\eta_i := \max_{\myvar{x} \in \myset{A}^i} \|P_i\myvar{e}_i\| $, then $\myvar{u}^* \in \myset{U}$ for all $\myvar{x} \in \cap_{i \in \myset{N}} (\myset{A}^i \cup \myset{C}^i)$.
\end{enumerate}
\end{proposition}
\begin{proof}
 (i) First, for all $\myvar{x} \in \myset{C}^i$, $c_i \geq 0$. We differentiate $h_i$ for $c_i \geq 0$ and $h_i \in [0, a]$: \rev{$\dot{h}_i$ $=$ $- \frac{1}{2}\gamma_i \myvar{e}_i^T P_i \bracketmat{cc}{\cos(x_3) & \sin(x_3)}^T u_{s_{p_i}}$ $=$ $\frac{1}{2}k_{p_i} \gamma_i^2 c_i \left( \myvar{e}_i^T P_i \bracketmat{cc}{\cos(x_3) & \sin(x_3)}^T \right)^2$ $\geq$ $0$.}
 We then differentiate $h_i$ for $h_i \in [-b, 0]$ for which $c_i \leq 0$, which yields $\dot{h}_i$ $=$ $-k \gamma_i^2 c_i \left( \myvar{e}^T P_i \bracketmat{cc}{\cos(x_3) & \sin(x_3)}^T \right)^2 \geq 0$. Since $\dot{h}_i$ $\geq$ $0$ on $\myset{A}^i$, we choose $\alpha(h) = \bar{\alpha}(h) \text{ if } h \geq 0, \alpha(h) = 0 \text{ if } h< 0 $, for any class-$\mathcal{K}$ function $\bar{\alpha}$, such that Properties \ref{prop:typeII} and \ref{prop:typeII robust} are satisfied and $\dot{h}_i \geq - \alpha(h_i)$ in $\myset{A}$. Thus it is clear that \eqref{eq:typeII ZCBF condition} holds, each $h_i$ is a Type-II ZCBF, and $\myvar{u}_{s_i}$ is locally Lipschitz continuous. Since each $\myset{A}^i$ has an empty intersection with $\myset{A}^j$ for all $j \in \myset{N}\setminus \{i\}$ and $P_i$ is positive definite such that $\nabla h_i \neq 0$ when $h_i = 0$, safety of each $\myset{C}^i$ follows from Theorem \ref{thm:Multiple ZCBF MIC}.
 
 (ii) If a) holds, then there is a $\myset{C}^i$, $i \in \myset{N}$ that is a compact forward invariant set such that $\myvar{x}(t) \in \myset{C}^i$ is bounded for all $t\geq 0$. If b) holds, $\myvar{x}(t)$ is bounded for all $t \geq 0$ as stated. Thus for either case a) or b), $\myvar{x}(t)$ is bounded and every $\myset{A}^i$ is compact for $i \in \myset{N}$ because each $P_i$ is positive-definite. Note that this holds regardless if $\gamma_i$ is $1$ or $-1$. Let $\myset{D}^i = \myset{C}^i \cup \myset{A}^i$, which is a connected (possibly non-compact) set, for each $i \in \myset{N}$. Now we investigate the case when $\dot{h}_i = 0$. Let $\Omega^i := \{ \myvar{x} \in \myset{D}^i\setminus \myset{C}^i: \dot{h}_i(\myvar{x}) = 0\}$, and $\dot{h}_i = 0$ when 1) $c_i = 0$, 2) $\myvar{e}_i = 0$ or 3) $\myvar{e}_i^T P_i \bracketmat{cc}{\cos(x_3) & \sin(x_3)}^T = 0$. In case 1), $c_i = 0$ only occurs when $\myvar{x} \in \myset{C}^i$ and so $\myvar{x} \notin \Omega^i$. In case 2), $\myvar{e}_i = 0$ does not occur in $\myset{A}^i$ by assumption and so the associated $\myvar{x}$ for which $\myvar{e}_i = 0$ is not in $\Omega^i$. 
 
In case 3) we can exclude the case when $\myvar{e}_i = 0$ since this is considered in case 2). Let $\myvar{\zeta}_i = P_i \myvar{e}_i$, for which $\myvar{\zeta}_i = (\zeta_{i_1}, \zeta_{i_2})$. We re-write case 3) as $\zeta_{i_1} \cos(x_3) + \zeta_{i_2} \sin(x_3) = 0$. First, we claim that $\dot{x}_1 = \dot{x}_2 = 0$ and $\dot{x}_3 \neq 0$ in $\Omega^i$. Since $\zeta_{i_1} \cos(x_3) + \zeta_{i_2} \sin(x_3) = 0$, it is clear from \eqref{eq:unicycle dynamics}, \eqref{eq:unicycle control} that $\dot{x}_1 = \dot{x}_2 = 0$. Now substitute $\dot{x}_3 = 0$ into \eqref{eq:unicycle dynamics}, which yields: $u_{s_{d_i}} = -k_{d_i} ( -\zeta_{i_i} \sin(x_3) + \zeta_{i_2} \cos(x_3)) = 0$. For $\dot{x}_3 = 0$, we need $-\zeta_{i_i}\sin(x_3) + \zeta_{i_2} \cos(x_3) = 0$. However apart from the case when $\myvar{e}_i = 0$, both $\zeta_{i_1} \cos(x_3) + \zeta_{i_2} \sin(x_3) = 0$ and $-\zeta_{i_i}\sin(x_3) + \zeta_{i_2} \cos(x_3) = 0$ cannot simultaneously be true for any $x_3 \in S^1$ (proof by contradiction and noting that $ \not\exists x_3$ s.t. $\sin(x_3) = \cos(x_3) = 0$). Thus $\dot{x}_3 \neq 0$ in $\Omega^i$. Suppose that a solution can stay identically in $\Omega^i$, i.e., if $\myvar{x}(\tau_1) \in \Omega^i$ for some $\tau_1 \in \mathbb{R}_{\geq 0}$, then $\myvar{x}(t) \in \Omega^i$ for all $t \geq \tau_1$. For $\myvar{x}$ to stay identically in $\Omega^i$, we require that $\dot{h} = 0$ for all $t \geq \tau_1$. Since $\myvardot{x} \neq 0$ in $\Omega^i$, the only way $\dot{h} = 0$ holds for all $t \geq \tau_1$ is if $\frac{d}{dt}[\zeta_{i_1} \cos(x_3) + \zeta_{i_2} \sin(x_3)] = 0$ for all $t\geq \tau_1$. Taking the derivative of $\zeta_{i_1} \cos(x_3) + \zeta_{i_2} \sin(x_3)$ and substituting $\dot{\zeta}_{i_1} = \dot{\zeta}_{i_2} = \dot{x}_1 = \dot{x_2} = 0$, yields: $(-\zeta_{i_1} \sin(x_3) + \zeta_{i_2} \cos(x_3) )\dot{x}_3= 0$.  Since $\dot{x}_3 \neq 0$ and there exists no $x_3 \in S^1$ such that $-\zeta_{i_1} \sin(x_3) + \zeta_{i_2} \cos(x_3) = 0$ and $\zeta_{i_1} \cos(x_3) + \zeta_{i_2} \sin(x_3) = 0$ hold simultaneously, there must exist some $\tau_2 > \tau_1$ such that $\myvar{x}(\tau_2) \notin \Omega$. Thus  no solution can stay identically in $\Omega^i$, and asymptotic stability of each $\myset{C}^i$ for $i \in \myset{N}$ follows from Theorem \ref{thm:typeII safety AS}.
 
 (iii) Since $\myset{A}^i$ is compact, $\eta_i$ is well defined, and $|c_i| \leq \max \{a_i, b_i\}$ for all $\myvar{x} \in \myset{A}^i$, $i \in \myset{N}$. It is clear that if \eqref{eq:unicycle gains} holds, then $\myvar{u}_{s_i} \in \myset{U}$ for all $\myvar{x} \in \myset{A}^i$, for all $i \in \myset{N}$. Since $\myset{U}$ is convex, then the proof follows from Theorem \ref{thm:Multiple ZCBF MIC}.
\end{proof}

\rev{We have presented several Type-II ZCBFs from Example \ref{ex:passivity1} and Proposition \ref{prop:unicycle}, which ensure safety and asymptotic stability to the safe set, but are \emph{not} ZCBFs as per Definition \ref{def:zcbf}. The proposed formulation generalizes the concept of ZCBFs, while retaining the desired properties of safety and robustness, yet also is able to handle multiple (Type-II) ZCBFs and input constraints simultaneously.}

\section{Numerical Results}

\begin{figure}
    \centering
    \includegraphics[scale=0.091]{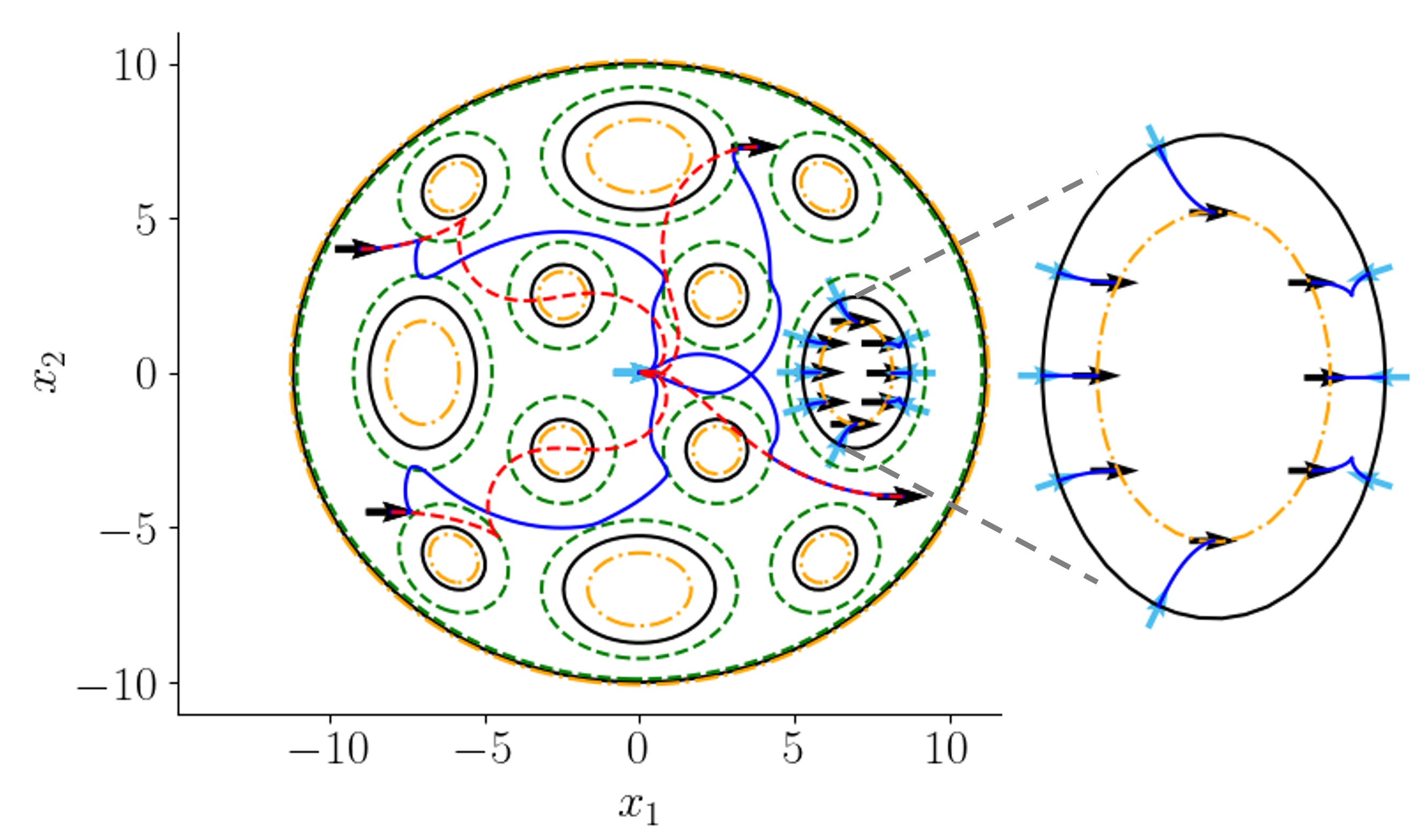}
    \caption{Simulation results for the proposed (blue solid curves) and nominal (red dashed curves) control. \rev{$\mysetbound{C}^i$, $\mysetbound{A}^i$ (\rev{w.r.t} $h_i = a_i$), and $\mysetbound{A}^i$ (\rev{w.r.t} $h_i = -b_i$) are depicted by solid black, dashed green, and dash-dotted orange curves, resp}. The initial and final configurations are depicted by black and sky-blue arrows, resp. }
    \label{fig:unicycle exp}
\end{figure}

Here \rev{the} goal is for a unicycle to navigate an obstacle-rich environment to reach $\myvar{x} = 0$. There are 12 obstacles plus a workspace boundary yielding $N = 13$ ellipsoidal constraints from \eqref{eq:unicycle constraint} (see Figure \ref{fig:unicycle exp}). We define each Type-II ZCBF as \eqref{eq:unicycle h} with $\myset{C}^i$ and $\myset{A}^i$ defined respectively by \eqref{eq:safe set i} and \eqref{eq:constraint annulus i}. The nominal controller is the stabilizing controller\cite{Aicardi1995}, $\myvar{u}_{nom} = (u_{nom_p}, u_{nom_d})$,  $u_{nom_p} = -k_r r \cos(\alpha)$,$u_{nom_d} = -k_a \alpha - \frac{k_r \sin(\alpha) \cos(\alpha) (\alpha - \theta)}{\alpha}$ where $r = \sqrt{x_1^2 + x_2^2}$, $\theta = \arctan(\frac{x_2}{x_1})$, $\alpha = x_3 - \theta$, and $k_r, k_a \in \mathbb{R}_{>0}$. To ensure $\myvar{u}_{nom} \in \myset{U}$ ($\myset{U}$ defined by \eqref{eq:unicycle input set}) with $\bar{u}_p = \bar{u}_d = 2.0$, $k_r, k_a$ are chosen such that $k_r \leq \frac{\bar{u}_p}{r(0)}$, $k_a \leq \frac{\bar{u}_d - k_r 1.5 \pi}{2 \pi}$. For each $\myset{C}^i$, $\myvar{u}_{s_i}$ is defined by \eqref{eq:unicycle control} with gains satisfying \eqref{eq:unicycle gains} such that $\myvar{u}_{s_i} \in \myset{U}$ and $\kappa(h_i) = -\frac{2}{a_i^3}h_i^3 + \frac{3}{a_i^2} h_i^2$.

\begin{figure}
    \centering
    \includegraphics[scale=0.45]{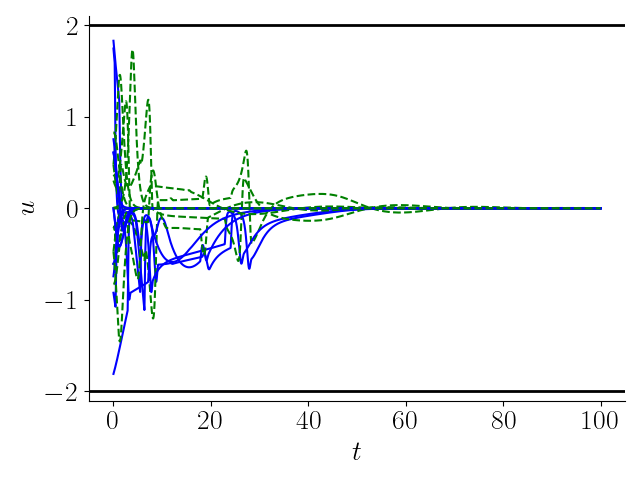}
    \caption{Input trajectory of proposed control. The blue solid curves,  green dashed curves, and solid black lines depict $u_p(t)$, $u_d(t)$, and the boundaries of $\myset{U}$, resp.}
    \label{fig:unicycle input}
\end{figure}

Figure \ref{fig:unicycle exp} shows the implementation of both the proposed control and the nominal control for various initial conditions. As expected, the trajectories associated with the proposed control (blue curves) avoid all obstacle regions, while converging to the origin. On the other hand, for the same initial conditions, the nominal control alone runs the unicycle through obstacle regions. To demonstrate the asymptotic stability, initial conditions were placed inside an obstacle region (zoomed part of Figure \ref{fig:unicycle exp}). For this case, we implemented the safety controller alone ($\myvar{u}_{nom} = 0$), for which the unicycle is pushed (backwards) outside the obstacle region. Figure \ref{fig:unicycle input} shows the control input trajectories for the proposed control, which satisfy the desired input constraint as dictated by Theorem \ref{thm:Multiple ZCBF MIC}.

\section{Conclusion}

We proposed a Type-II ZCBF for ensuring forward invariance and robustness of a constraint set, \rev{which} is more general than the original ZCBF formulation. \rev{We also proposed} a new control design that accommodates multiple Type-II ZCBFs with non-intersecting constraint set boundaries, while respecting input constraints. The proposed approach was applied to the classical unicycle system. Future work will address non-intersecting constraint set boundaries.

\bibliographystyle{IEEEtran}
\bibliography{IEEEabrv,typeIIZCBF}

\end{document}